\documentclass[12pt,preprint]{aastex}
\def\alwaysmath#1{\ifmmode{#1}\else{$#1$}\fi}
\def\msun{\alwaysmath{{M}_{\odot}}}

\def\arcsec{\hbox{$^{\prime\prime}$}}

\def\pula{PSR~J1911$-$5958A~}
\def\compula{COM~J1911$-$5958A~}

\def\ltsima{$\; \buildrel < \over \sim \;$}
\def\gtsima{$\; \buildrel > \over \sim \;$}
\def\lsim{\lower.5ex\hbox{\ltsima}}
\def\gsim{\lower.5ex\hbox{\gtsima}}
\def\lapp{\ifmmode\stackrel{<}{_{\sim}}\else$\stackrel{<}{_{\sim}}$\fi}
\def\gapp{\ifmmode\stackrel{>}{_{\sim}}\else$\stackrel{<}{_{\sim}}$\fi}

\slugcomment{In press in the ApJ Letters}

\begin{document}

\title{The Helium White Dwarf orbiting the Millisecond
Pulsar in the halo of the Globular Cluster NGC6752
\footnote{Based on
observations collected by using the Very Large Telescope,
at the European Southern Observatory, Cerro Paranal, Chile, 
within the observing programme 71.D-0232.}}

\author{Francesco R. Ferraro\altaffilmark{2},
Andrea Possenti\altaffilmark{3,4},
Elena Sabbi\altaffilmark{2},  
Nichi D'Amico\altaffilmark{3,5}}
\medskip

\affil{\altaffilmark{2}{Dipartimento di Astronomia Universit\`a 
di Bologna, via Ranzani 1, I--40127 Bologna, Italy,
ferraro@bo.astro.it}} 
\affil{\altaffilmark{3}{Osservatorio Astronomico di Cagliari,
Loc. Poggio dei Pini, Strada 54, I--09012 Capoterra, Italy}}
\affil{\altaffilmark{4}{Osservatorio Astronomico di Bologna, 
via Ranzani 1, I--40127 Bologna, Italy}}
\affil{\altaffilmark{5}{Dipartimento di Fisica Universit\`a di Cagliari,
Cittadella Universitaria, I-09042 Monserrato, Italy}}

\begin{abstract}
We have used deep high-resolution multiband images taken at the ESO
{\it Very Large Telescope} to identify the optical binary companion to
the millisecond pulsar (\pula$\!\!$)  located in the halo of the
Galactic Globular Cluster NGC6752. The object turns out to be a blue
star whose position in the Color Magnitude Diagram is consistent with
the cooling sequence of a low mass ($M\sim 0.17-0.20M_{\odot}$), low
metallicity Helium white dwarf (He-WD) at the cluster distance.  This
is the second He-WD which has been found to orbit a millisecond
pulsar in GGCs.  Curiously both objects have been found to lie on the
same mass He-WD cooling sequence.  The anomalous position of \pula
with respect to the globular cluster center ($\sim 6\arcmin$)
suggested that this system has recently ($\lapp 1$ Gyr) been ejected
from the cluster core as the result of a strong dynamical
interaction. The data presented here allows to constrain the cooling
age of the companion within a fairly narrow range ($\sim 1.2-2.8 $
Gyr), therefore suggesting that such dynamical encounter must have acted
on an already recycled millisecond pulsar.
\end{abstract}
 
 
\section{Introduction} 
\label{sec:intro}

\pula has been discovered on 1999 October 17 during a search for
Millisecond Pulsar (MSPs) in Galactic Globular Clusters (GGCs) in
progress at the Parkes Radiotelescope \citep{damico01}.  It is a
binary millisecond pulsar with a spin period of 3.26 ms, an orbital
period of 0.84 days and very low eccentricity ($e<10^{-5}$).  Precise
celestial coordinates ($RA= 19^{\rm h}\, 11^{\rm m}\, 42\fs756$, $
DEC=-59\arcdeg\, 58\arcmin\, 26\farcs91$) have been recently obtained
for this source from pulsar timing observations \citep{damico02}.
This position is far away ($\sim 6\arcmin$) from the cluster optical
center: indeed \pula is the more off-centered pulsar among the sample
of 44 MSPs whose position in the respective cluster is known, and it
suggests that this object might be the result of strong interactions
occurred in the cluster core.

\citet{colpi02} explored a number of possibilities for the peculiar
location of\\ \pula$\!\!$: a careful analysis led to discard the
hypothesis of a primordial binary (born either in the halo or in the
core of the cluster) and to reject also the possibility of a 3-body
scattering or exchange event off core stars. Hence they conjectured
that a more massive target \citep[either a binary
or a single intermediate-mass black hole, see
also][]{colpi03a,colpi03b} could have provided the necessary thrust to
propel \pula into its current halo orbit at an acceptable event rate.

Moreover, beside \pula$\!\!$, NGC6752 hosts also the second most
displaced MSP ever seen in a globular (PSR J1911$-$6000C) and at least
two (out the three) MSPs located in the cluster center \citep[PSR-B
and PSR-E, see Figure 2 in][]{ferraro03} experience strong
acceleration, implying an unusually high central mass to light ratio
$(M/L\gsim 6)$ \citep{damico02,ferraro03}. \citet{ferraro03} showed
that NGC6752 is a dynamically evolved cluster probably undergoing a
post-core-collapse bounce and investigated scenarios for
simultaneously explaining both the anomalous acceleration of PSR-B and
PSR-E and the ejection of \pula and PSR J1911$-$6000C, concluding that
the existence of a binary black hole of intermediate mass
\citep{colpi02} could be a viable possibility.

Optical detection of the companion to a millisecond pulsar in a
globular cluster proved particularly helpful in assessing the origin
and the evolution of the binary, besides supporting its cluster
membership \citep[see, e.g.,][]{edmonds01,ferraro01a,edmonds02}.
In fact, unlike the systems in the galactic field, the age,
metallicity, extinction, distance and hence intrinsic luminosity and
radius can be estimated from the parent cluster parameters. Hence, we
undertook a systematic program \citep{ferraro01a,ferraro01b}
devoted to the optical identification of MSPs companion in GGCs; as a
part of this project in this Letter we present the identification of
the optical counterpart to the PSR J1911-5958A
companion.
   
\section{Observations and data analysis}
\label{sec:obs}
 
The photometric data presented here consist of a set of high
resolution images obtained by using the camera FORS1 mounted at the
{\it ANTU} Unit Telescope 1 (UT1) of the Very Large Telescope (VLT) at
ESO on Cerro Paranal (Chile) on three nigths in March, April and May
2003. All the observations have been performed in the High Resolution
(HR) mode of FORS1. In this configuration the plate-scale is
$0\farcs1/pixel$ and the FORS1 $2048\times2048$ pixel$^2$ array has a
global field of view of $3\farcm4 \times 3\farcm4$.  The data comprise
eight 220\,s $V$-band exposures, five 360\,s $B$-band exposures and
three 1500\,s $U$-band exposures, roughly centered on the \pula
(hereafter MSP-A) nominal position. All the observations were
performed in service mode under good seeing conditions
(FWHM$=0\farcs5-0\farcs7$).

A sub-image of $800\times 800$ pixels centered on the nominal position
of the MSP-A has been extracted from the original frames and carefully
analized.  All the reductions have been carried out using ROMAFOT
\citep{buonanno83}, a package specifically developed to perform
accurate photometry in crowded fields, allowing for a visual
inspection of the quality of the PSF-fitting procedure.  In order to
optimize the search for faint objects a {\it median}-combined image in
each band has been obtained and the search procedure has been
performed on the deepest combined image.  Then the masks with the star
positions have been adapted to each image and the PSF-fitting
procedure performed on each individual frame separately.  The
resulting instrumental magnitudes have been transformed to a common
photometric system and then averaged. Thus, a final catalogue with the
coordinates and the average instrumental magnitude in each filter has
been compiled for all the stars identified in the considered
sub-image.  Photometric calibration of the instrumental magnitudes (in
the $B$ and $U$ band) was obtained using four photometric standard
stars \citep{landolt92} secured under photometric conditions. Since no
accurate calibration of the V filter was possible with the data
secured in service, we calibrated the $V$ magnitude by using $\sim
200$ stars in common with the $B,V$ catalog recently published by
\citet{ferraro03}. The stars in common between the two catalogs
permitted also an independent check of the calibration obtained in the
$B$ band, displaying agreement within a few hundredths of magnitude.

The stars identified in the considered sub-image have been reported to
an absolute astrometric system by using the 200 stars in common with
the \citet{ferraro03} catalog (already astrometized).  The details
of the procedure adopted to derive the astrometric solution are
described in other papers \citep{ferraro01b,ferraro03}. In short, the
new astrometric {\it Guide Star Catalog} ($GSCII$) was used to search
for astrometric standard stars lying in the considered field of view (FoV).
At the end of the procedure, the rms residuals were of the order of
$\sim 0\farcs 3$ both in RA and Dec and we assume this value as
representative of the astrometric accuracy.  
 
\section{Results}
   
Figure~1 shows the $U,U-V$ and $B,B-V$ Color Magnitude Diagram (CMD)
for the stars ({\it large filled circles}) identified in the
$80\arcsec\times 80\arcsec$ FORS1 sub-image considered here. Most of the stars
trace a clean and well defined main sequence (MS) spanning almost 8
mags in the $U$ band reaching $U\sim 26$.  Only a few sparce objects,
showing a blue excess, are visible on the left side of the MS. The 
most extreme blue objects in the CMDs are CO white dwarfs (WDs): once
the MS is matched, they nicely overlaps the position of a CO-WDs
population (shown as {\it small empty circles} in Fig.~1)
observed in this cluster by \citet{renzini96} and
\citet[][2003b, in preparation]{ferraro97}. In particular, the colors and the
luminosities of the three CO-WDs found here and of the previously
observed population agree with the theoretical cooling sequence for
$0.5 M_{\odot}$ hydrogen rich WDs \citep{wood95} (drawn as {\it heavy
dashed line} in Fig.~1).
   
On the basis of the accurate astrometric positions obtained from the
photometric catalog, we identified a blue object (hereafter \compula$\!\!$)
lying at only $0\farcs 1$ from the nominal position of \pula$\!\!$.
Finding chart (in the $B$-band filter) for \\ 
\compula is shown in
Fig.~2. Only a few objects ($\sim 10$) are lying in the region of the CMD 
to the left of the MS: considering the FoV of the sub-image, there is a
very small probability ($\lsim 5\times 10^{-3}$) of detecting one of
the blue objects just by chance in a circular aperture of radius $\sim
0\farcs 3$ (corresponding to the uncertainty in the relative radio-optical
astrometry) centered on the pulsar position.
     
The position of \compula is marked as a {\it large empty square} in
the CMDs in Fig.~1, whereas absolute coordinates and magnitudes are
listed in Table~1. Taking into account the expected variability of
this object and the global uncertainty in the calibration of the
photometric zero-point, we conservatively adopted an overall
uncertainty in the reported magnitudes of 0.15 mag.  While we were
writing this Letter a not yet refereed paper by Bassa et al. appeared
on the web (astro-ph/0307340) also presenting the optical
identification of this object. Though the identification is consistent
with that presented here, it can be considered, at most, a preliminary
detection since the object is nearly at the detection limit in the
exposures utilized in that paper. The deep VLT observations presented
here allow us to measure the object with a quite high S/N ratio
($70-120$) and to derive the photometric properties of the star with
much higher accuracy.

The location of \compula in the CMDs ($\sim 1.5$ mag bluer than the MS
in the diagram) excludes it is a CO-WD whilst it resembles $U_{opt}$,
the companion to PSR J0024--7203U in 47 Tuc \citep{edmonds01},
which was suggested \citep{edmonds01} to be a low mass Helium core WD
(He-WD) from the comparison with theoretical models by
\citet{serenelli01}. However, the He-WD tracks used by
\citet{edmonds01} were computed for progenitors with high (solar)
metallicity ($Z=0.02$).
 
Here we have applied the same procedure taking advantage of a new set
of tracks specifically computed \citep{serenelli02} for globular
cluster applications, i.e. assuming progenitors with low
metallicities. In particular, we used the cooling tracks at $Z=0.001,$
which are the closest to the cluster metallicity
($[Fe/H]=-1.43\pm0.04$) as recently derived by \citet{gratton03}. The
cooling sequences for two He-WD masses (0.172 and 0.197 \msun,
respectively) are over-plotted in Fig.~1 ({\it light dashed lines}).
The models have been drawn by adopting a distance modulus $(m-M)_0
=13.13$ \citep[from the homogeneous distance scale for 61 GGCs derived
by][]{ferraro99} and a reddening $E(B-V)=0.04$
\citep{harris96,gratton03}.  Note that the distance modulus adopted
here is consistent both with the estimate obtained from the WD cooling
sequence ($(m-M)_0 =13.05\pm 0.1$) by \citet{renzini96} and with the
most recent determination ($(m-M)_0 =13.12\pm 0.08$) derived from MS
fitting by \citet{gratton03}.
 
Inspection of Fig.~1 reveals that luminosity and color of \compula
well agree with its being a He-WD of mass in the range
$0.17-0.20~\msun.$ In particular, from the $Z=0.001$ models of masses
$0.172\msun$ and $0.197\msun$ by \citet{serenelli02}, we derive the
following estimates for the properties of \compula$\!\!$: a temperature in
the interval $T_{\rm eff}= 10,000-12,000~K$, a gravity $\log
g=6.12-6.38$, a luminosity $L=0.03-0.04~L_{\odot}$, a radius
$R=3-4\times 10^9$ cm and a cooling age in the range $1.2-2.8$ Gyr. 

\section{Discussion}

We have identified the optical companion to the millisecond pulsar
\pula located at 3.3 half mass radii from the center of the
globular cluster NGC6752. $U, B$ and $V$ magnitudes (and 
related colors) of the optical source are compatible with
its being a Helium white dwarf hosted in the globular's halo.

From the pulsar mass-function \citep[0.0026887 \msun,][]{damico02}, it
results a minimum companion mass (corresponding to a system seen
edge-on, i.e. with orbital inclination $i=90^\circ$) $M_{\rm
COM}=0.185~\msun$ for a pulsar mass of $1.35~\msun$
\citep{thorsett99}.  Given the upper limit ($M_{\rm COM}\lapp
0.2~\msun$) of the range of masses of \compula inferred from the
cooling tracks of \citet{serenelli02}, we can constrain the
inclination $i$ to be larger than $\gapp 70^\circ$. Adopting a larger 
value for the pulsar mass would result in a even larger lower limit
for the orbital plane inclination. 

Dynamical friction can rapidly drive back to the cluster core a object
of total mass of order $1.6~\msun$ moving on a highly eccentric orbit
in the cluster potential. In particular, ejection of \pula from the
core to the outskirts of NGC6752 (implying a radial orbit) cannot be
occurred more than $\tau_{\rm df}=0.7-1$ Gyr ago \citep{colpi02}. The
cooling age of \compula is longer than $\tau_{\rm df},$ implying that
any dynamical event responsible for the ejection probably acted on an
already recycled millisecond pulsar. Hence, the optical identification
of \compula tends to exclude {\it (i)} the scenarios in which the
recycling process occurred after (or was triggered by) the dynamical
encounter which propelled the pulsar in the cluster halo (but see the
discussion in Sigurdsson, 2003) and {\it (ii)} any kinds of
interactions imparting a significant eccentricity to the binary pulsar
(having excluded a recent mass transfer phase, the circularization of
the system at the present level, $e\lsim 10^{-5},$ would require a
time much longer than $\tau_{\rm df}$). In particular, among the
proposed black hole hypotheses \citep{colpi03a,colpi03b} a binary
black hole of mass in the range 10-200 \msun~seems the most plausible
candidate.

Since the first discovery of an optical counterpart to a binary MSP
companion in GGC \citep[$U_{opt}$, the companion to PSR~J0024$-$7203U
in 47 Tuc;][]{edmonds01}, the zoo of the optical MSP counterparts in
GGCs is rapidly increasing: a surprisingly bright object
(COM~J1740$-$5340, either a subgiant or a heated MS) has been
discovered \citep{ferraro01b} in NGC6397, and a very faint source
(W29, the companion to PSR~J0024$-$7204W) has been found by
\citet{edmonds02} in 47 Tuc. Moreover, an additional potential MSP
companion (W34 in 47 Tuc) has been recently discussed by
\citet{edmonds03}. Fig.~3 shows a comparison of the photometric
properties of the available optical identifications of MSP companions
hosted in GGCs. The MS stars of NGC6752, the models of CO-WD and two
He-WD cooling sequences by \citet{serenelli02} are also plotted in
Fig.~3 as reference.  Two among the five sources seem really peculiar:
i.e., the bright object in NGC6397 (which is as luminous as the turn
off stars and shows quite red colours) and the faint W29 in 47 Tuc,
which is also too red to be a He-WD \citep{edmonds02}.  Bassa et
al. (2003) suggested a similiarity between the photometric properties
of \compula and W29 in 47 Tuc.  As can be seen from Figure 3, \compula
is significantly brighter and bluer than W29 and it turns out to be
more similar to $U_{opt}$. Indeed, $U_{opt}$ and \compula are found to
lie nearly on the same mass He-WD cooling sequence and W34 in 47 Tuc
curiously shares the same photometric properties of \compula. Indeed,
if confirmed as a MSP companion, W34 would be the third He-WD
companion orbiting a MSP in GGCs roughly located on the same-mass
cooling sequence.  If further supported by additional cases, this
evidence could confirm that a low mass $\sim 0.15-0.2~\msun$ He-WD
orbiting a MSP is the favoured system generated by the recycling
process of MSPs not only in the Galactic field \citep{hp98} but also
in GGCs \citep{rasio00}. 
  
\compula is relatively bright and will allow detailed follow-up
observations: both optical time series photometry and spectroscopy of
this object are planned at the ESO VLT, in order to determine the
parameters of the He-WD companion, the mass ratio of the two
components of the binary and eventually the mass of the neutron star.
     
\acknowledgements{
We warmly thank Elena Valenti for assistance during the astrometry
procedure and Leandro Althaus for pointed out the updated He-WD
models. The authors wish to thank the Referee for helpful suggestions
in revising the manuscript.  The financial support of the Agenzia
Spaziale Italiana (ASI) is kindly acknowledged. A.P and
N.D'A. received financial support from the Italian Space Agency (ASI)
and the Italian Minister of Research (MIUR) under {\it Cofin 2001}
national program.}

\clearpage
  
\begin{figure} 
\plotone{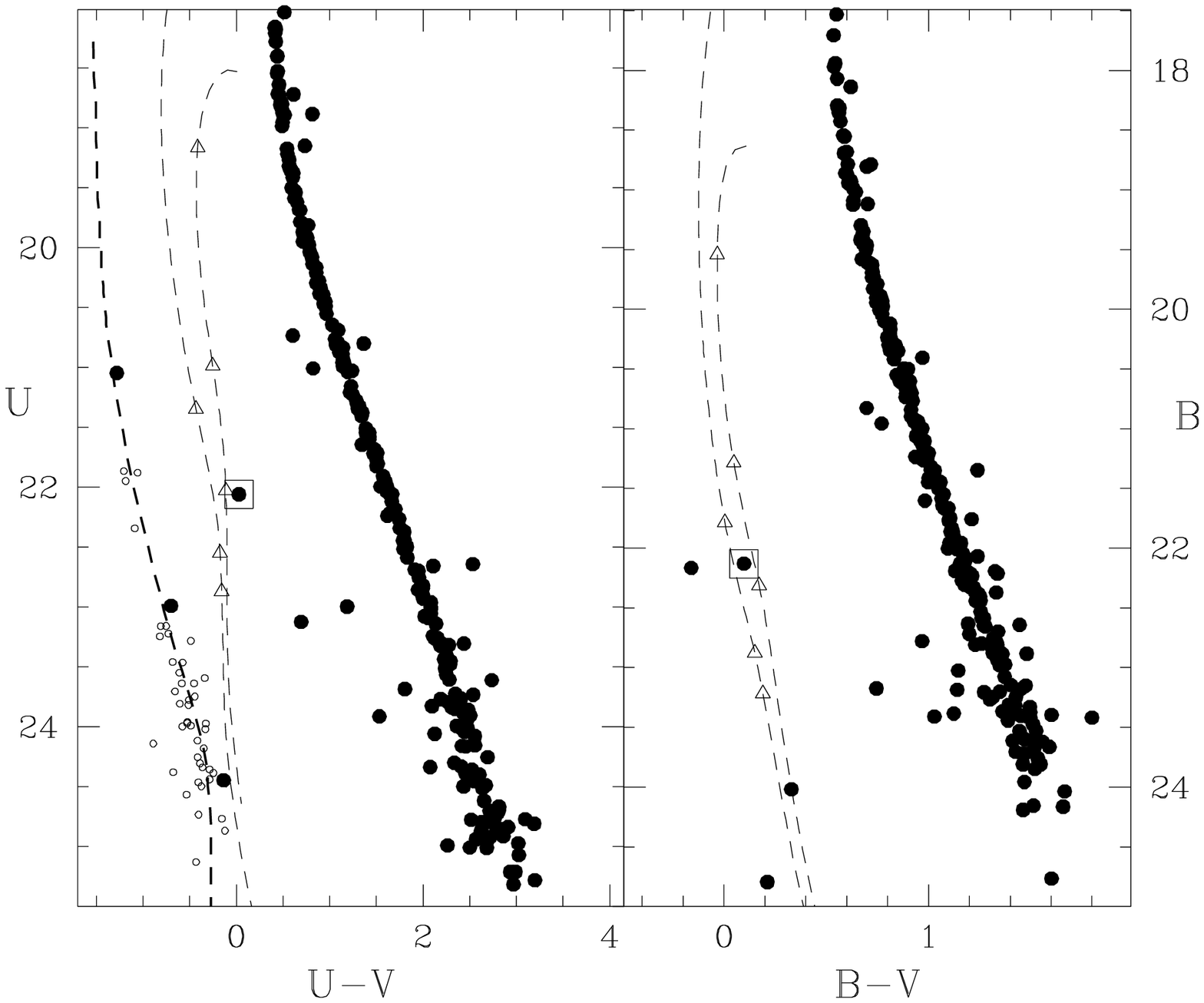} 
\caption{
($(U,U-V)$) and ($(B,B-V)$) CMDs for the stars identified in a region
of $80\arcsec \times80\arcsec$ centered at the nominal position of the
\pula. The optical counterpart to the pulsar companion
(\compula$\!\!$) is marked with a {\it large empty square}.  The {\it
heavy dashed line} is the CO-WD cooling sequence from \citet{wood95};
the two {\it light dashed lines} are the cooling tracks for He WD
masses 0.197 and 0.172 $M_{\odot}$ (the lowest mass model is the
reddest one) from \citet{serenelli02}. The {\it small triangles} along
the two tracks labels different ages (1,2 and 3 Gyr, respectively).
The CO-WD population observed in this cluster by Renzini et al. (1996)
and Ferraro et al. (1997, 2003b, in preparation) is also plotted as {\it
small open circles} in the left panel.
\label{fig:CMDs}}
\end{figure}
  
 \clearpage 

\begin{figure} 
\plotone{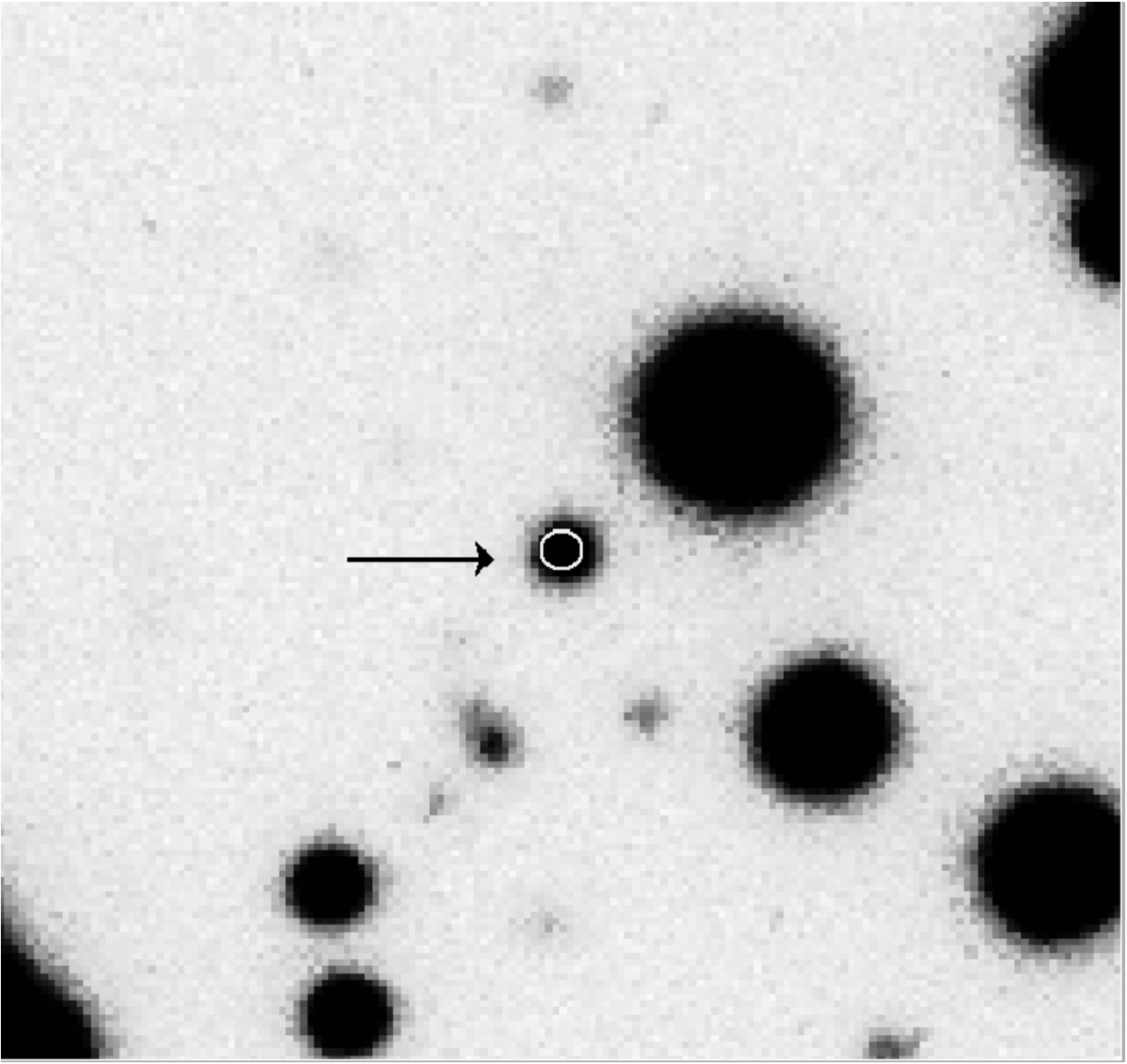} 
\caption{Finding chart for \compula showing the median-combined B-band
image. The region covers $15\arcsec \times15\arcsec$, also plotted is
the $1\sigma$ error circle ($0\farcs 3$) for the absolute astrometric
positioning of the optical image. North is up and East is to the left.
\label{fig:chart}}
\end{figure}

\clearpage
 
\begin{figure}
\plotone{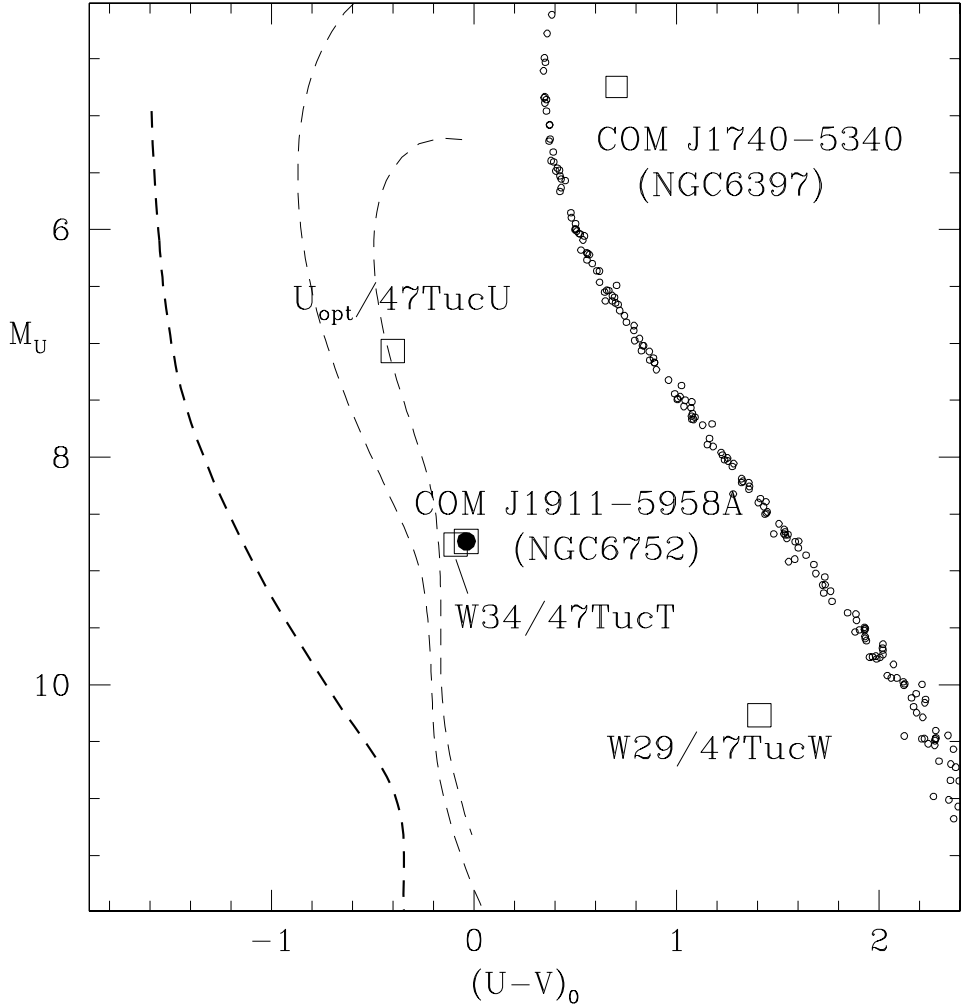} 
\caption{Optical companions to MSP detected in GGCs in the
($M_U,(U-V)_0$) absolute plane: \compula is plotted as in Figure 1,
while the other objects with a {\it large empty square}.  As in Figure
1, we present the cooling tracks for He-WD from Serenelli et
al. (2002) and the CO-WD cooling sequence from Wood (1995). Main
sequence stars of NGC6752 are also plotted ({\it small open circles})
as reference.  Note that $U_{opt}$ and \compula appear to lie along
the same He-WD cooling sequence. Also W34 in 47 Tuc (if confirmed to
be a MSP companion) shares the same photometric properties of
\compula$\!\!$. In order to place in the plot the 3 objects belonging to 47
Tuc, the distance modulus ($(m-M)_0=13.27$) and the reddening
($E(B-V)=0.055$) of this cluster have been adopted according to
Zoccali et al. (2001); the approximate position of COM J1740$-$5340 
(companion to PSR J1740$-$5340 in NGC6397) in the absolute plane has
been obtained by shifting the CMD of Figure 1 of Ferraro et
al. (2001b) to match the main sequence of NGC6752.
\label{fig:allMSPcomp}}
\end{figure}

\clearpage
 
\begin{deluxetable}{ccccccc}
\scriptsize
\tablewidth{15cm}
\label{tab:uve}
\tablecaption{Photometric data and position for \compula}
\tablehead{
\colhead{} &
\colhead{} &
\colhead{$ V$} &
\colhead{$ B$} &
\colhead{ $U$} &
\colhead{$\alpha_{\rm J2000}$} &
\colhead{$\delta_{\rm J2000}$}  }
\startdata
\compula & & 22.03 & 22.13 & 22.06 &
 $19^{\rm h}\, 11^{\rm m}\, 42\fs743$  &
$ -59\arcdeg\, 58\arcmin\, 26\farcs85$  \\ 
\enddata
\end{deluxetable}
   
\end{document}